\documentstyle[12pt,epsf]{article}
\renewcommand{\baselinestretch}{1.0}

\newcommand{\be}{\begin{equation}}
\newcommand{\ee}{\end{equation}}
\begin{document}
\topmargin 0pt
\oddsidemargin=-0.4truecm
\evensidemargin=-0.4truecm
\renewcommand{\thefootnote}{\fnsymbol{footnote}}

\newpage
\setcounter{page}{1}
\begin{titlepage}     
\vspace*{-2.0cm}
\begin{flushright}
\vspace*{-0.2cm}
\end{flushright}
\vspace*{0.5cm}

\begin{center}
{\Large \bf Searches for sterile  component with  solar neutrinos and KamLAND}
\vspace{0.5cm}

{P. C. de Holanda$^{1,2}$ and  A. Yu. Smirnov$^{1,3}$\\
\vspace*{0.2cm}
{\em (1) The Abdus Salam International Centre for Theoretical Physics,  
I-34100 Trieste, Italy }\\
{\em (2) Instituto de F\'\i sica Gleb Wataghin - UNICAMP, 
13083-970 Campinas SP, Brazil}\\
{\em (3) Institute for Nuclear Research of Russian Academy 
of Sciences, Moscow 117312, Russia}

}
\end{center}

\begin{abstract}
Possible mixing of the active and sterile neutrinos has been considered 
both in the single  $\Delta m^2$ approximation and in the case of 
more than one $\Delta m^2$.   We perform global fit of the available solar neutrino data 
with free  boron neutrino flux in the single  $\Delta m^2$ context. 
The best fit value corresponds to zero  fraction  of sterile component
$\eta = 0$. We get the upper bounds:  $\eta < 0.26 (0.64)$ at $1\sigma$ ($3\sigma$). 
Due to degeneracy of parameters no one individual experiment restricts $\eta$. 
The bound appears as an interplay of the SNO and Gallium as well as SuperKamiokande data. 
Future measurements of the NC/CC ratio at SNO can strengthen  the bound down to 
$\eta < 0.5$ ($3\sigma$). If KamLAND confirms the LMA solution with its best fit point 
a combined analysis of the KamLAND and solar neutrino results will lead to  $\eta 
< 0.19 (0.56)$ at $1\sigma$ ($3\sigma$). 
We find that existence of sterile neutrino can explain  the  
intermediate value of  suppression of   the KamLAND event rate: 
$R_{KL} \sim 0.75 - 0.90$ in the case when more than one $\Delta m^2$ is involved. 

\end{abstract}

\end{titlepage}
\renewcommand{\thefootnote}{\arabic{footnote}}
\setcounter{footnote}{0}
\renewcommand{\baselinestretch}{0.9}

\section{Introduction} 

Sterile neutrino, $\nu_s$,  is a {\it panacea} of various problems which 
appear from time to time in  neutrino physics.  
Oscillation interpretation of the LSND result~\cite{lsndresult} 
in the context of schemes which  describe  the solar and atmospheric neutrino
problems~\cite{4faminterpr} is the long term motivation for existence of $\nu_s$. 

There are, however,  other motivations to search for the sterile component.  
It might happen that $\nu_s$ is the
right-handed component of neutrino which turns out to be  light (or
massless) in the context of the see-saw mechanism, as a consequence of
certain symmetry~\cite{seesaw}.  Such a symmetry ($L_e - L_{\mu} - L_{\tau}$
as in~\cite{seesaw} ) can also be responsible for large or maximal lepton
mixing.  Sterile neutrino can be a component of multiplet of the
extended gauge group, {\it e.g.}, $E_6$ \cite{e6}. 
Sterile neutrino - light singlet fermion - can originate from completely
different sector of theory.  Many extensions of the standard model
predict existence of singlet fermions: $\nu_s$ can be a mirror
neutrino~\cite{mirror}, goldstino in SUSY~\cite{susy}, modulino of the
superstring theories~\cite{superstring}, bulk fermions related to existence of
extra dimensions~\cite{extradimension}, etc..

Neutrinos are unique particles, since  
only they can mix with singlets of the Standard Model. 
This  mixing can lead  to coherent effects which provide high sensitivity 
to existence of $\nu_s$.  The mixing  terms  as 
small as $(10^{-10} - 10^{-6})$ eV can  produce the observable effects 
and in the case of supernova neutrinos the masses  can be even smaller. 
In a sense,  neutrino mixing is a window to physics beyond the SM.

Another motivation to search for and restrict sterile component is
that interpretation of certain results can be 
changed strongly if one admits even very small admixture  of sterile neutrinos. 
This concerns with the neutrinoless double beta decay, 
CP-violation, generation of maximal/large flavor mixing \cite{balaji}, etc.. 

So,  we  deal  with long term program of searches for sterile neutrinos 
and improvements of the bounds on mixing of sterile neutrinos 
with all possible masses. 

In this paper we will consider in details searches for sterile
component in the solar neutrino flux.  First detailed discussion of
signatures of the sterile neutrinos have been done in
~\cite{BG93,BG95,BGtaup} where it was marked, in particular, that studies of the
spectrum distortion can reveal existence of $\nu_s$.

There were a number of fits of the solar neutrino data in terms of
pure $\nu_e \rightarrow \nu_s$ sterile conversion, but 
recent SNO results~\cite{sno-nc,sno-dn} exclude pure $\nu_e \rightarrow \nu_s$ 
conversion at high confidence level. Still partial transformation of
$\nu_e$ in $\nu_s$ is possible. The effects, and allowed fraction
of sterility depend on specific scheme of mixing. Most of recent
studies have been done in a single $\Delta m^2$ context: 
according to which  the electron neutrino mixes  with the state  being  the combination of the
active ($\nu_{\mu}$ and $\nu_{\tau}$) and sterile neutrino. Fraction
of sterility is described by the parameter $\eta$.  In~\cite{barger}
it was marked that $\eta$ is weakly restricted provided that the
original boron neutrino flux is substantially larger than the SSM
flux.  Global fit of the solar neutrino data~\cite{concha} gives 
$\eta < 0.25$ ($0.6$) at $1\sigma$ ($3\sigma$) level. 
If the boron neutrino flux is fixed according to the SSM 
with corresponding uncertainties the bound becomes stronger at 
the $3\sigma$  level:  $\eta < 0.5$ ~\cite{jose}.

In this paper we will consider effects of sterile neutrinos and 
bounds on sterility  from the existing and future experiments. 
In particular, we study how combined analysis of the KamLAND and 
solar neutrino data can improve the bound. We generalize the analysis 
for the case when more that  one $\Delta m^2$ is  relevant. 
We discuss how presence of sterile component can change predictions
for KamLAND rate.  

The paper is organized as follows. In section 2 we consider the mixing 
the sterile neutrino in a single $\Delta m^2$ context.  
In section 3 we perform  global
fit of all solar neutrino data, and put bounds on the sterile
fraction $\eta$. In section 3.2 we identify the data sensitive to $\eta$. 
In section 4 we study  how future experiments can improve the bound. 
We consider  precision $NC$ measurements  
at  SNO (section 4.1) and  combined analysis of KamLAND and solar
neutrino data (section 4.2).  
In section 5 we consider the $\nu_s$ mixing in the case when more than 
one $\Delta m^2$ is relevant. We discuss how the presence of sterile component 
can change predictions for the  KamLAND experiment. 
We present  our conclusions in section 6.

\section{Mixing of sterile neutrinos}  

In what follows we will consider the case of one sterile neutrino. 
In general, fourth neutrino may have an arbitrary mass and 
sterile component can mix with all three active neutrinos. 
So, the scheme  will have 4 new real parameters: the mass $m_4$ and 
3 mixing angles. Situation can, however, be simplified if one takes into
account existing bounds on mixing of active neutrinos.
We will consider the 4-neutrino schemes which explain  
the solar and atmospheric neutrino data. 
We assume that two mass eigenstates, 
$\nu_1$ and $\nu_2$,  are splited by the solar 
$\Delta m^2 < several \times 10^{-4}$ eV$^2$.  
We will call them the ``solar pair". Then the state $\nu_3$, is splited from 
the solar pair or from $\nu_4$ by the atmospheric mass split: 
$\Delta m^2_{atm} \sim (2 - 3) \cdot 10^{-3}$ eV$^2$. 
We define the mixing matrix, $||U_{\alpha i}||$ as the unitary matrix which connects 
the flavor and the mass eigenstates: $\nu_{\alpha}  = \sum_i U_{\alpha i} \nu_i$, 
$\alpha = e, \mu, \tau, s$, $i = 1, 2, 3, 4$.

\subsection{Single $\Delta m^2$ case} 
\label{sec:singledm}

As far as solar neutrinos are concerned, mixing of the  electron neutrinos
plays crucial role. 

Let us first assume that  $U_{e3} = U_{e4} =0$,  so that the electron flavor is
distributed only in the ``solar pair of states":  
In fact, there is  a strong upper bound on $U_{e3}$ from CHOOZ experiment~\cite{chooz},  
and strong bound on $U_{e4}$ from BUGEY experiment~\cite{BUGEY} 
if  corresponding $\Delta m^2$ is large.

Since only $U_{e1}$ and $U_{e2}$ are non-zero the 
orthogonality conditions for $\nu_e$ and other neutrinos, 
$\nu_{\mu},~ \nu_{\tau},~ \nu_s$ can be
written as 
\be
U_{e1}^{\dagger} U_{\alpha 1} + U_{e2}^{\dagger} U_{\alpha 2} = 0~, ~~~~~
\alpha = s, \mu, \tau~.  
\label{orth}
\ee
From these equations we get immediately: 
\be
\frac{U_{\mu 2}}{U_{\mu 1}} = \frac{U_{\tau 2}}{U_{\tau 1}} = 
\frac{U_{s2}}{U_{s1}} = - \frac{U_{e2}^{\dagger}}{U_{e1}^{\dagger}}. 
\label{propo}
\ee  
That is, all non-electron flavor components  enter  
$\nu_1$ and $\nu_2$ in  the same combination. 
Indeed, let $\nu_x$ be the combination of the non-electron neutrinos 
which mixes with $\nu_e$ in $\nu_1$: 
\be 
\nu_1 = \cos \theta_{12} \nu_e - \sin \theta_{12} \nu_x, ~~~~~ 
\nu_x = \frac{1}{\sqrt{1 - U_{e1}^2}} 
\sum_{\alpha} U_{\alpha 1} \nu_{\alpha}~,   
\label{nu1}
\ee
where  $\sin \theta_{12} = \sqrt{1 - U_{e1}^2}$. Similarly, let 
$\nu_y$ be the combination of the non-electron neutrinos
which mixes with $\nu_e$ in $\nu_2$:  
\be
\nu_2 = \sin \theta_{12} \nu_e + \cos \theta_{12} \nu_y, ~~~~~
\nu_y = \frac{1}{\sqrt{1 - U_{e2}^2}}
\sum_{\alpha} U_{\alpha 2} \nu_{\alpha}. 
\label{nu2}
\ee
Then according to (\ref{propo}), we have:  
\be
\nu_x = \nu_y.
\label{xy}
\ee
Thus, the scheme is reduced exactly to the  two two neutrino case:  
$\nu_e$ mixes with $\nu_x$ in the states $\nu_1$ and  $\nu_2$ 
and the mixing angle equals $\theta_{12}$. 

The state $\nu_x$ can be written as 
\be
\nu_x = \sqrt{1 - \eta} \nu_a +  \sqrt{\eta} \nu_s,   
\label{nux}
\ee
where $\nu_a$ is the combination of active (non-electron) neutrinos, 
$\nu_{\mu}$ and $\nu_{\tau}$:
\be
\nu_a=\frac{1}{\sqrt{1-U_{s1}^2}}(U_{\mu 1}\nu_{\mu} + U_{\tau 1}\nu_\tau). 
\label{aaa}
\ee
So,  $\sqrt{\eta}$ describes
admixture of  sterile neutrino in the  state $\nu_x$ to which 
$\nu_e$ can be transformed. 
According to (\ref{nu1}), (\ref{nu2}) and (\ref{nux}), 
admixtures of the sterile component in $\nu_1$ and $\nu_2$ equal 
\be
U_{s1} = \sqrt{\eta} \sin \theta_{12}, ~~~ 
U_{s2} = \sqrt{\eta} \cos \theta_{12},   
\label{stU}
\ee 
and consequently,  
\be
|U_{s1}|^2 + |U_{s2}|^2 =  \eta. 
\label{norm}
\ee       
That is,  $\eta$ 
gives total fraction of the sterile component in the solar pair.

We have proven that if the electron neutrino is distributed in two
mass eigenstates only, then independently of other mixings and masses, 
the effect of sterile component in solar neutrinos is described by one
parameter only, which is the total amount of sterile component in these
two mass eigenstates. 

\subsection{Conversion probabilities and observable fluxes}

Let us  consider  the oscillation effects. We introduce    
\be 
P_{ee} = P_{ee}(\Delta m^2_{12}, \theta_{12}, V_{ex})  
\ee 
-  the $\nu_e$ survival probability in the system of mixed neutrinos
$\nu_e  - \nu_x$.
Here $V_{ex} = V_e - V_{\mu}(1 - \eta)$ is the effective potential.   
According to  (\ref{nu1}) and (\ref{nux}) the transition
probabilities of  the electron neutrino to the sterile, $P_{es}$,   
and  active, $P_{ea}$, components in terms of $P_{ee}$ equal    
\begin{eqnarray}
P_{es} &=& \eta(1 -   P_{ee}) \nonumber\\
P_{ea} &=& (1 -  \eta) (1 -   P_{ee}).
\end{eqnarray}
Using these probabilities we can write 
the fluxes of neutrinos which determine the rates of events of different types.  
Let us introduce $f_B$, the original boron neutrino flux in units 
of the SSM flux, :   
\be
f_B \equiv \frac{F_B}{F_B^{SSM}}~.  
\ee
Here the SSM  boron neutrino flux is taken to be 
$F^{SSM}_B = 5.05 \cdot 10^{6}$ cm$^{-2}$ s$^{-1}$.  
Then the elastic scattering events (ES) (detected by 
SuperKamiokande  and SNO), the neutral current 
(NC) and charged current (CC) event rates at SNO are determined by the
following fluxes:  
\begin{eqnarray}  
\Phi_{\rm CC}  &=& f_B P_{ee} \, , 
\label{eq:RSNOCC} \\
\Phi_{\rm NC}  &=& f_B [P_{ee}+(1-\eta)(1-P_{ee})] \, ,
\label{eq:RSNONC}\\  
\Phi_{\rm ES}& = & f_B [P_{ee} + r(1-\eta)(1 - P_{ee})] \,,  
\label{eq:RSK} 
\end{eqnarray}  
where $r=0.16$ is the ratio of the $\nu_{\mu,\tau}\,e^-$ and
$\nu_{e}\,e^-$ elastic scattering cross sections.  
The equations (\ref{eq:RSNONC}) depend on three 
parameters,  $f_B$,   $P_{ee}$ and $\eta$  via two combinations
\be
x \equiv f_B P_{ee}, ~~~ y \equiv  f_B (1- \eta)(1 - P_{ee}).  
\label{twocomb}
\ee
In terms of variables $x$ and $y$, the fluxes can be rewritten as
\be
\Phi_{ES} = x + r y, ~~~ \Phi_{CC} = x, ~~~ \Phi_{NC} = x +  y.
\label{fluxeq}
\ee
Excluding $x$ and $y$  we get relation between the  fluxes \cite{barger}:
\be
\Phi_{ES} = \Phi_{CC} + r(\Phi_{NC} - \Phi_{CC}) = \Phi_{CC}(1 - r) +
r \Phi_{NC}
\label{sum}
\ee
which has a simple interpretation 
(see the first equality):  the flux measured in the neutrino-electron
scattering equals the flux of electron neutrinos at the detector
plus the flux of non-electron active neutrinos,  $\Phi_{NC} - \Phi_{CC}$,  
suppressed by $r$.  
The equality (\ref{sum}) is well satisfied in the experiment.

The equalities (\ref{fluxeq})  imply the {\it degeneracy of parameters} \cite{barger}: 
different values of the  original parameters $f_B$, $\eta$, $P_{ee}$ 
lead to the   same observables provided that they keep to be constant 
the combinations $x$ and $y$. 
In particular, changes of $\eta$ can be compensated 
by corresponding variations of 
$P_{ee}$ and $f_B$, so that the combinations and therefore the 
observables are not changed. Clearly this is possible 
if $P_{ee}$ does not depend on the energy or time. 
Therefore the energy and the time variations  of the probability 
break the degeneracy. In other words, time variations and distortion of the spectrum are, 
in general, sensitive to the fraction of sterile neutrinos. 
Since  no significant distortion or variations 
have been found experimentally, the present data have no strong sensitivity to the
fraction $\eta$ as we  will see from exact calculations in the 
sect.  3.

\section{Global fit and bounds on fraction of sterile component}
\label{glo}

\subsection{Global analysis}

We have performed a global analysis of all available solar 
neutrino data taking into account possible presence of the 
sterile component. We follow the procedure described 
in our previous publication~\cite{us:msw} where details can be found,  
and here we summarize the  main ingredients of the analysis.

We use the same data sample as in  \cite{us:msw}, which consists of 

\noindent
- three total rates: 1) the $Ar$-production rate, $Q_{Ar}$,  from  Homestake~\cite{Cl}, 2) the
$Ge-$production rate,  $Q_{Ge}$ from  SAGE \cite{sage} and 3) the combined
$Ge-$production rate from GALLEX and GNO \cite{gno}; 

\noindent
- 44 data points from the zenith-spectra measured by Super-Kamiokande \cite{SK} 
during 1496 days of operation;

\noindent
- 38 day-night spectral points from SNO \cite{sno-nc,sno-dn}. 

All solar  neutrino fluxes (but the boron neutrino flux) are taken
according to SSM BP2000 \cite{ssm}. 
We use the boron neutrino flux as free parameter. 
For the $hep-$neutrino flux we take fixed value  
$F_{hep} = 9.3 \times 10^{3}$ cm$^{-2}$ s$^{-1}$ \cite{ssm,hepfl} .

In contrast to~\cite{us:msw}, the neutrino scheme includes now 
mixing with sterile component which is described by the parameter $\eta$~(\ref{nux}). 
So, there are three oscillation parameters:  
the mass squared difference, $\Delta m^2$, and two mixings: $\tan^2\theta$ and $\eta$. 
Consequently, in the free boron neutrino flux fit 
we have four parameters: $\Delta m^2$, $\tan^2\theta$, $\eta$
and $f_B$, and therefore  there are 81(data points) - 4 = 77 d.o.f..

We perform the $\chi^2$ test of the oscillation solution by
calculating
\be
\chi^2_{global} = \chi^2_{rate} +   \chi^2_{SK} + \chi^2_{SNO}, 
\label{chi-def}
\ee
where $\chi^2_{rate}$, $\chi^2_{SK}$  and  $\chi^2_{SNO}$
are the contributions
from the total rates, the Super-Kamiokande zenith  spectra
and the SNO day and night spectra correspondingly. Each of 
the entries in (\ref{chi-def}) is the function of four
parameters ($\Delta m^2$, $\tan^2\theta$, $f_B$, $\eta$), 
in particular,   
\be
\chi^2_{global} = 
\chi^2_{global}(\Delta m^2, \tan^2\theta, f_B, \eta).   
\ee 

We perform the global analysis for various {\it fixed} values
of $\eta$. We first find the best fit points for a given $\eta$ minimizing
$\chi^2_{global}$ with respect to $\Delta m^2$, $\tan^2 \theta$ and $f_B$.
This gives $\chi^2_{min}(\eta)$.  
The results are shown in the Table 1.  
The best fit corresponds to zero value of sterile fraction and
$f_B = 1.05$. So, the  absolute minimum is characterized by
$\chi^2_{min}(0)$. 
According to the Table 1, $\chi^2$ increases
with $\eta$, and moreover, the increase is fast for $\eta > 0.3$. 
$\Delta \chi^2 = \chi^2_{min}(0.4) - \chi^2_{min}(0) = 2.4$;  
for $\eta = 0.6$ this difference is $\Delta \chi^2 = 7.5$.

\begin{figure}[ht]
\centering\leavevmode
\epsfxsize=.8\hsize
\epsfbox{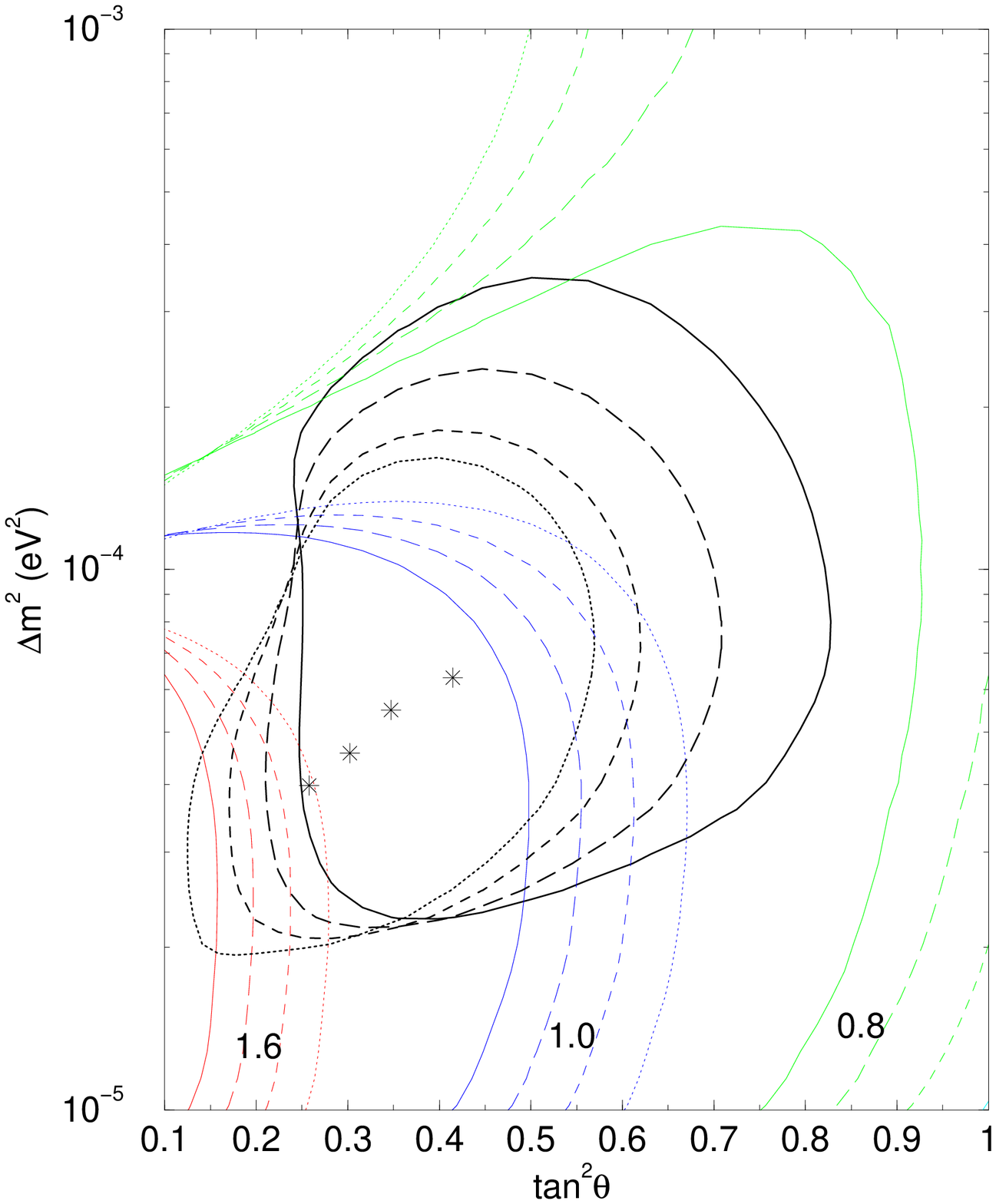}
\caption{The best fit points and 3$\sigma$ allowed regions of oscillation parameters 
for different values of $\eta$: $\eta$ = 0.0 (solid line), 0.2
(long-dashed 
line), 0.4 (dashed line) and 0.6 (dotted line).  Boron neutrino
flux is taken as free parameters. Also shown lines of constant $f_B$
which minimize $\chi^2$.}
\label{fig:chi2all}
\end{figure}

Then we construct the $3\sigma$ regions in the
$\Delta m^2 - \tan^2 \theta$ plane (Fig.~\ref{fig:chi2all})  minimizing
$\chi^2 (\Delta m^2, \tan^2 \theta, f_B)$  with respect to $f_B$:
we denote the  corresponding minimal value of $\chi^2$ by
$\chi^2_f (\Delta m^2, \tan^2 \theta)$. Then  $3\sigma$ contours
are determined by  
$\chi^2_f (\Delta m^2, \tan^2 \theta)-\chi^2_{min}(\eta)=11.73$, 
where $\chi^2_{min}(\eta)$ is the minimum $\chi^2$ for a given 
value of $\eta$.
We have found also lines of constant values of $f_B$ which minimize
$\chi^2$ for a given $\eta$.

According to Fig. \ref{fig:chi2all} and the Table 1, with increase of $\eta$ the
allowed region shifts to smaller values of $\tan^2 \theta$ and  $\Delta
m^2$.  At the same time $f_B$ and $\chi^2_{min}$ increase.

\begin{table}
\begin{center}
\begin{tabular}{ccccc}
\hline $\eta$         & 0.0  & 0.2  & 0.4  & 0.6    \\
\hline $\chi^2$       & 65.1 & 65.5 & 67.5 & 72.6   \\ 
\hline $\Delta m^2 \, (\times 10^{-5})$ & 6.3&5.5&4.6&4.0 \\ 
\hline $\tan^2\theta$ & 0.41 & 0.35 & 0.30 & 0.26   \\ 
\hline $f_B$         & 1.046& 1.188& 1.356& 1.570   \\ 
\hline
\end{tabular}
\label{tab:chi2}
\caption{Minimum $\chi^2$ for different values of $\eta$. Also
presented are the values of parameters $\Delta m^2$, $\tan^2\theta$ and $\eta$ that
minimize $\chi^2$.}
\end{center}
\end{table}

These features  can be well understood using Eqs. 
(\ref{eq:RSK} - \ref{twocomb}). 
Increase of $\eta$ [decrease of $(1 -  \eta)$] can be 
compensated by  increase of $f_B$ and decrease of $P_{ee}$. 
Since $P_{ee} \sim \sin^2 \theta$, the decrease of $P_{ee}$ implies 
decrease of mixing angle. With decrease of mixing the distortion of the 
spectrum (mainly the turn up of the probability at low energies)  
becomes stronger. This leads to shift  of the allowed region to smaller
values of $\Delta m^2$ (shift of the spectrum from the adiabatic edge).

Let us consider breaking of parameter degeneracy. 
Taking $P_{ee} \approx \sin^2 \theta$  in  expressions 
(\ref{twocomb}) and using results  of the Table 1  
we find values of parameters $x(\eta)$ and $y(\eta)$  
for different $\eta$ in the best fit points (see Table \ref{xyy}).  
As follows from the Table~\ref{xyy} in the best fit points 
 the parameters $x(\eta)$ and $y(\eta)$ are not invariant under changes of $\eta$: 
$x$ slightly increases (by $5\%$) whereas $y$ decreases significantly: 
almost by $50\%$. 
Another way to see these variations is to keep $x = \Phi_{CC} = constant$ 
and to exclude $P_{ee}$ from the combination $y$:  
\be
y = (1 - \eta) (f_B - \Phi_{CC}). 
\ee
Using $\Phi_{CC} = 0.32$ we find $y(\eta)$ shown in the third line of the Table~\ref{xyy}.  
Thus, the invariance of $x$ and  $y$ (and therefore a degeneracy) 
is broken which leads the dependence of 
$\chi^2_{min}$ on $\eta$ and therefore to 
sensitivity of the analysis to the presence  of sterile neutrinos. 
\begin{table}
\begin{center}
\begin{tabular}{ccccc}
\hline 
$\eta$    & 0.0  & 0.2  & 0.4  & 0.6    \\
\hline 
$x$       & 0.304 & 0.308 & 0.313 & 0.320   \\ 
\hline 
$y$       & 0.742  & 0.703 & 0.623 &  0.492 \\ 
\hline 
$y(x=const)$ & 0.726  & 0.694 & 0.622 & 0.500   \\ 
\hline 
\end{tabular}
\label{xyy}
\caption{Values of parameters $x$ and $y$ in the best fit points 
of global analysis for different values of $\eta$.}
\end{center}
\end{table}

\begin{figure}[ht]
\centering\leavevmode
\epsfxsize=.8\hsize
\epsfbox{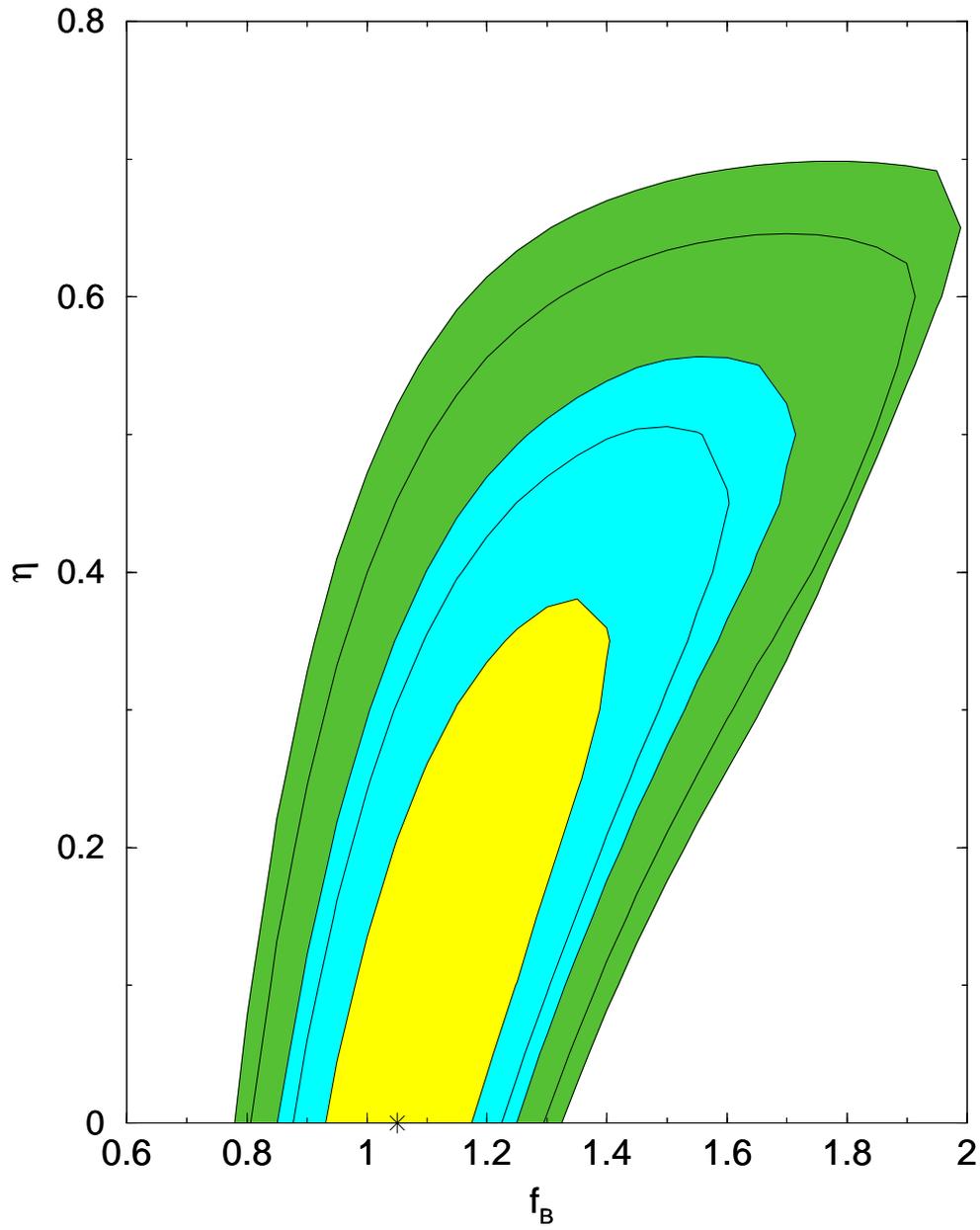}
\caption{The allowed regions in $f_B - \eta$ plane, at  1$\sigma$, 90\%,
95\%, 99\% and 3$\sigma$ C.L.. The best fit point (marked by star) is at $\eta = 0$. }
\label{fig:etafb}
\end{figure}

For fixed values of $\eta$ and $f_B$ we minimize $\chi^2$
with respect to $\Delta m^2, \tan^2 \theta$ which gives 
$\chi^2_{\Delta, \theta} = \chi^2_{\Delta, \theta} (\eta, f_B)$. Using
this function we construct contours of constant  confidence level in $\eta, f_B$ plane
(fig.~\ref{fig:etafb}).  
From this figure we get an upper limit for the sterile fraction 
\begin{equation}
\eta < 0.38 \,(0.70),\,\,\,1\sigma\,\,(3\sigma) 
\label{etalimits1}
\end{equation}
The boron neutrino flux which correspond  to maximal allowed  value of $\eta$ equals  
$f_B = 1.25$  ($1\sigma$) and $f_B = 1.68$ ($3\sigma$). 

We have performed minimization of $\chi^2$ with respect to $\Delta m^2, \tan^2 \theta$ 
and $f_B$, thus finding $\chi^2_{\Delta, \theta, f_B} = 
\chi^2 (\eta)$.  Then the function $\chi^2 (\eta)$  allows to put  the bounds for 1 d.o.f.: 
\begin{equation}
\eta < 0.26 \,(0.64),\,\,\,1\sigma\,\,(3\sigma) 
\label{etalimits2}
\end{equation}
These results  are very similar  to results obtained in \cite{concha,jose}. 
In particular, it was found~\cite{concha} that at $1\sigma$:  
$\eta_{max} =  0.25$  for $f_B \sim 1.25$,  and at  $3\sigma$:  
$\eta_{max} =  0.6$  for $f_B \sim 1.75$. 

Imposing the SSM bound on the neutrino flux 
further strengthen the bound on ``sterility". 
It was found in~\cite{jose}: $\eta < 0.25$  at the $1\sigma$ and 
$\eta < 0.5$  at the $3\sigma$ level.  
Notice that in fact $1 \sigma$ bound is unchanged since 
at this level the required boron neutrino flux is within 
$1\sigma$ SSM prediction. 
In contrast, the limit becomes stronger at the $3\sigma$ level, 
where substantially larger than in SSM flux is needed.

For the SSM value of the boron neutrino flux, $f_B = 1$, we get 
from the fig.~\ref{fig:etafb} the following bounds on the sterile
fraction: $\eta <  0.14$ ($1\sigma$), 
$\eta < 0.30$ ($2\sigma$),  $\eta < 0.47$ ($3\sigma$).

\subsection{Who does not like sterile neutrino?}

Let us identify observables which are  sensitive to  $\eta$. 
As we have discussed in sect.~\ref{sec:singledm} 
these observables should 
contain information about the time variations  or/and  the 
energy dependence of the conversion effect. 
These dependences  remove the degeneracy and therefore restrict $\eta$. 

No statistically significant time variations have been found
although some indications of the Day- Night asymmetry, $A_{ND}$,  exist~\cite{SK,sno-dn}.
Let us consider if  $A_{ND}$ can  restrict  the admixture of sterile neutrino.
The high energy neutrino data can be  described  by  
the $\nu_e$-survival probability during the day and during the night:  
$P_{ee}^D$, $P_{ee}^N$, or equivalently, by $\Delta P_{ee} \equiv  P_{ee}^N - P_{ee}^D$ and
$\bar{P}_{ee} \equiv  (P_{ee}^N + P_{ee}^D)/2$. 
Consequently, in Eq. (\ref{eq:RSNOCC} - \ref{eq:RSK})
we should substitute $P_{ee} \rightarrow \bar{P}_{ee}$ and add
to analysis an  additional observable:
\be
A_{ND} = \frac{\Delta P_{ee}}{\bar{P}_{ee}}.
\label{dnas}
\ee
In the LMA region the average suppression is determined basically by
mixing angle: $\bar{P}_{ee} \sim \sin^2 \theta$, whereas the DN asymmetry
depends mainly on $\Delta m^2$. This can be seen from 
lines of constant ratio $CC/NC \sim P_{ee}$ and lines of constant $A_{ND}$
which are nearly orthogonal each other 
(see figs. 5 and 6 from \cite{us:msw}).  Therefore $\Delta P_{ee}$ and
$\bar{P}_{ee}$ do not correlate and  can be considered as independent
fit parameters. For a given $\bar{P}_{ee}$, the difference  $\Delta P_{ee}$ 
can change in the wide range, and the data on the asymmetry can be easily fitted.  
So, adding experimental information about the Day and
Night asymmetry does not help to resolve degeneracy and therefore to improve
the bound on $\eta$. 
Precise measurements of the zenith angle dependence  of signal 
will change this conclusion.

No statistically significant distortion of the energy spectrum is found 
for $E > 5$ MeV, which supports the degeneracy problem.
The distortion (in experiments sensitive to $E > 5$ MeV)  
is expected for higher values of $\Delta  m^2$ 
and small mixing.  However, substantial distortion is expected over 
wider detectable energy range
which includes also the $pp-$ neutrinos. 
Therefore, the interplay of the high energy data and results from the low energy experiments 
should  break the degeneracy.   

To illustrate this effect, let us describe the distortion by two (for
simplicity) different values of probabilities: at high energies,
$P_{ee}^{H}$, and at low energies, $P_{ee}^L$ ($E < 0.5$ MeV).  The
signal in the Gallium experiment is  determined essentially by
$P_{ee}^L$: so that the $Ge$-production rate equals $Q_{Ge}
\propto P_{ee}^L$, and obviously, no substantial dependence on $f_B$
appears. 
In contrast to the Day-Night asymmetry case, the probabilities
$P_{ee}^L$ and $P_{ee}^{H}$ are strongly (anti) correlated. Indeed, in
the range of $pp$-neutrinos:
\be
P_{ee}^L \sim 1 - 0.5 \sin^2 2\theta,
\ee
and since $P_{ee}^{H} \approx \sin^2 \theta$ we get  
\be
P_{ee}^L \sim 1 - 2 P_{ee}^{H}(1 -  P_{ee}^{H}).
\label{problowenu}
\ee
With decrease of $P_{ee}^{H}$ the low energy probability
$P_{ee}^L$ increases. Consequently, the predicted value 
of the $Ge$-production rate increases. 
Thus,  combination  of the  Gallium and SNO  data should break the degeneracy  
of parameters. 

Now let us describe results of exact  numerical calculations. 

1). We have performed the $\chi^2$ analysis of the SNO data only. 
The best fit points in the oscillation plane as well as 
$f_B$ which minimize $\chi^2_{SNO}(\eta)$  are given for different values 
of $\eta$ in the Table~\ref{chi2sno}.
Notice that with increase of  $\eta$  the quality of the fit does not
change: $\chi^2$ even slightly decreases. So, as far as SNO data alone are
concerned, no bound on $\eta$  appears. Variations of $\eta$ can be
completely compensated by changes of $P_{ee}$ and $f_B$. The day-night asymmetry
has low statistical significance.

\begin{table}
\begin{center}
\begin{tabular}{ccccc}
\hline $\eta$         & 0.0  & 0.2  & 0.4  & 0.6    \\
\hline $\chi^2$       & 25.6 & 25.4 & 25.3 & 25.1   \\ 
\hline $\Delta m^2 \, (\times 10^{-5})$ & 4.5&4.5&4.0&3.6 \\ 
\hline $\tan^2\theta$ & 0.45 & 0.35 & 0.28 & 0.18   \\ 
\hline $f_B$         & 1.039& 1.201& 1.432& 2.013   \\ 
\hline
\end{tabular}
\label{chi2sno}
\caption{Minimum $\chi^2_{SNO}$ for different values of $\eta$. Also
presented are  values of parameters $\Delta m^2$, $\tan^2\theta$ and $\eta$ that
minimize the $\chi^2_{SNO}$.}
\end{center}
\end{table}

\begin{figure}[ht]
\centering\leavevmode
\epsfxsize=.8\hsize
\epsfbox{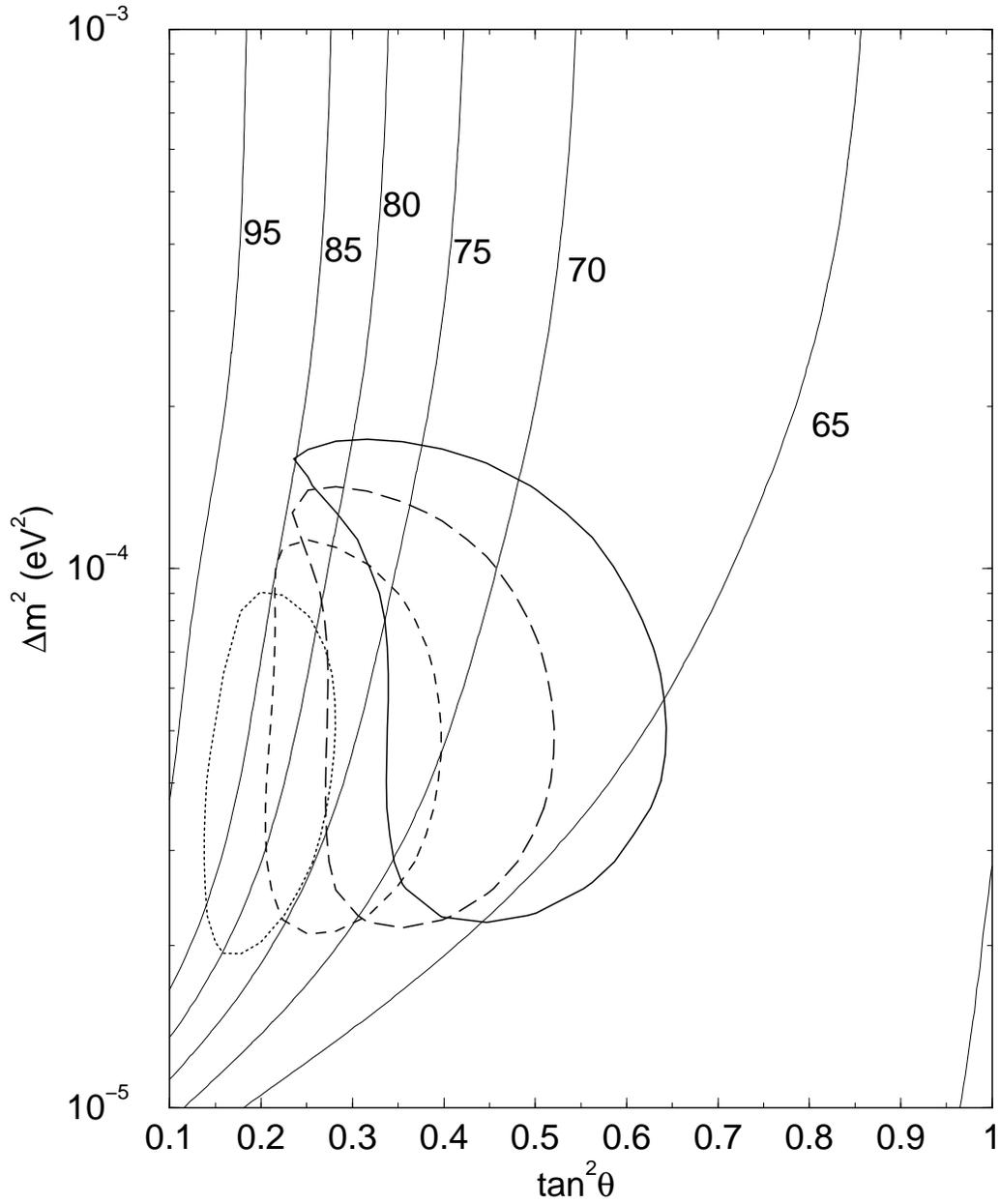}
\caption{The 1$\sigma$ allowed region from the analysis of the  SNO data only 
for $\eta=0$ (straight line), 0.2 (long dashed), 0.4
(dashed) and 0.6 (dotted). Also shown are lines of constant $Q_{Ge}$.}
\label{fig:fbsno}
\end{figure}

The SNO experiment has higher sensitivity to the  $NC$-
events. For this reason the allowed region provided
by SNO is restricted in $\tan^2\theta$ from both sides. 
Increase of  $\eta$ has the effect of moving the allowed region to
smaller values of $\tan^2\theta$, as can be seen in fig.~\ref{fig:fbsno}. 

Indeed, to keep the combinations  (\ref{twocomb}) unchanged with increases of $\eta$ one needs 
(i) to diminish $P_{ee}$, which implies decrease of $\theta$ since 
in the LMA region $P_{ee} \sim \sin^2 \theta$, (ii) to decrease  $\Delta m^2$
to avoid distortion (turn up) of the spectrum  at small energies 
and (iii) to  increase $f_B$ to compensate decrease of $P_{ee}$. 
 
2). In fig.~\ref{fig:fbsno} we show also lines of constant $Ge$-production rate.
The rate increases with decrease of mixing. According to the figure, 
in the SNO allowed region for $\eta = 0.6$  the rate $Q_{Ge} > 75$ SNU  
and  in the best fit point $Q_{Ge} = 83$ SNU. 
For  $\eta = 0.4$  we get $Q_{Ge} = 75$ SNU in the best fit point.  
The combined results from SAGE,  Gallex and GNO experiments is 
$Q_{Ge} = (70 \pm 4)$ SNU. 
So, the  Gallium results prevent from further shift of the allowed region to smaller 
$\theta_{12}$ and therefore forbid larger $\eta$.  
To illustrate the role of Gallium experiments, 
we find the bounds on $\eta$ that can be derived by taking only SNO and Gallium  data:
\begin{equation}
\eta < 0.43 \,(0.78),\,\,\,1\sigma\,\,(3\sigma),
\label{eta-snoG}
\end{equation}
which are weaker than those obtained by taking all the 
data~(\ref{etalimits2}).

In principle, further more precise measurements of the 
Ge-production rate could improve the bound on the sterile component. \\

\begin{figure}[ht]
\centering\leavevmode
\epsfxsize=.8\hsize
\epsfbox{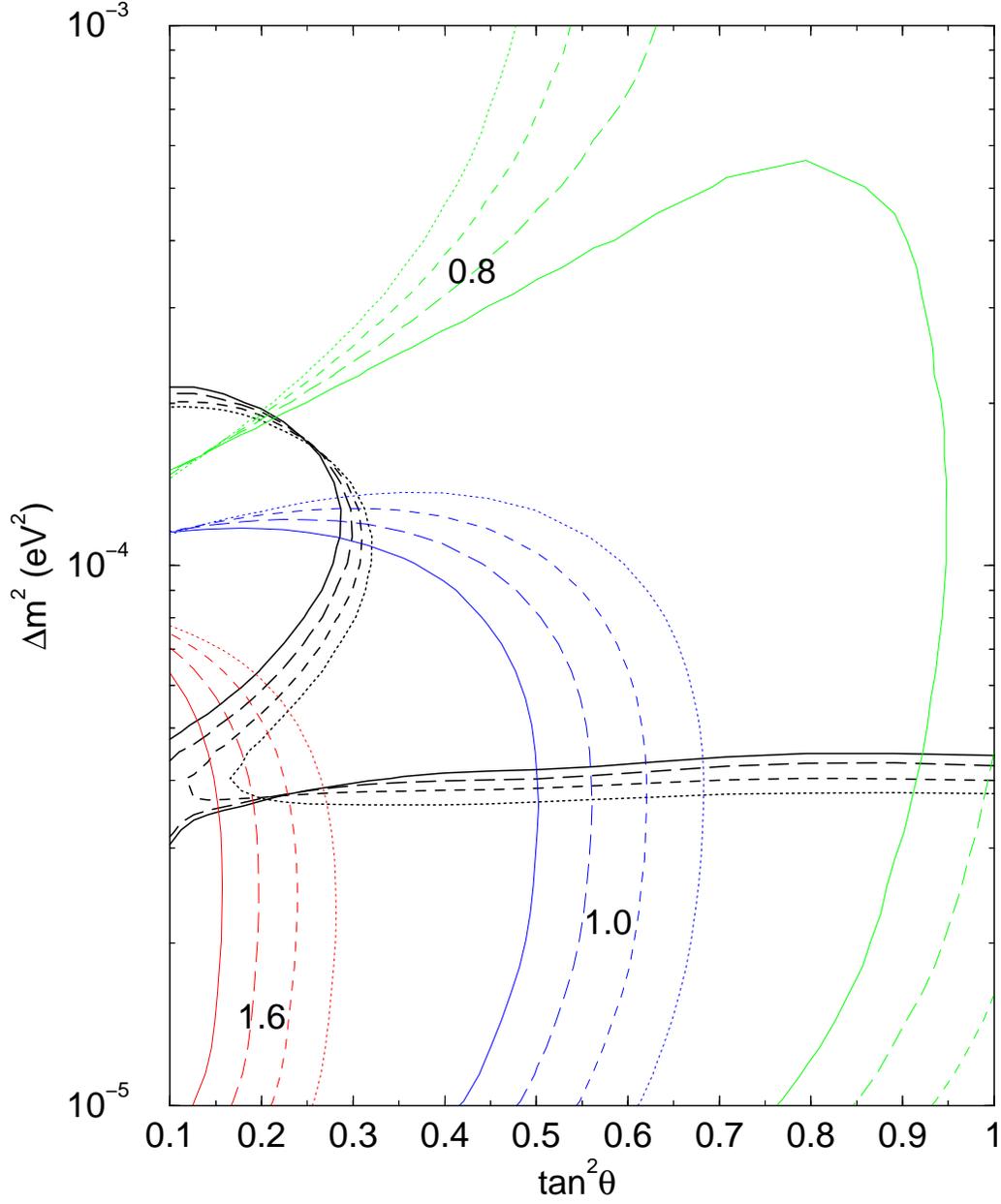}
\caption{The 1$\sigma$ allowed region (right upper parts)
from the analysis of the SuperKamiokande data for different values of $\eta$.  
Also shown  are lines of constant $f_B$ which minimize
$\chi^2$ fit of the SK data for different  $\eta$.} 
\label{fig:fbsk}
\end{figure}

3). Let us comment on the impact of SuperKamiokande. 
In fig.~\ref{fig:fbsk} we show result of analysis of the SuperKamiokande data only 
for different values of $\eta$.  The excluded (at $1 \sigma$) regions 
in the oscillation parameter space  
are due to the effects of spectrum distortion and the 
Earth $\nu_e$ regeneration. They  are only slightly modified by the 
presence of sterile component.
With increase of $\eta$ the excluded region shifts to larger mixing
angles. This is related to the fact that with increase of sterile fraction 
the distortion of the spectrum increases due to decrease of the damping effect 
related to active neutrinos.   The exclusion due  to  earth regeneration becomes weaker.  
Since for active-sterile system  the effective matter potential is smaller   
the regeneration region shifts to smaller $\Delta m^2$. 
Correspondingly, lower bound of the allowed region shifts to smaller $\Delta m^2$. 

In the allowed region, where distortion and time variations are small  
the  data can be reproduced for  different values of  
$\eta$ by adjusting $f_B$ and $P_{ee}$ (see  (\ref{eq:RSK})).

In fig.~\ref{fig:fbsk} we show also $f_B$ that gives the
best fit to the Super-Kamiokande data. 
With increase of $\eta$ the lines $f_B = const$ shift to larger
$\tan^2 \theta$  and $\Delta m^2$ (where the survival probability is larger)
to compensate decrease of muon and tau neutrino contribution.
Large values of $\eta$ are allowed, when  the  boron neutrino 
flux is much larger than the SSM flux.   
 
Similarly to the Gallium experiments, SuperKamiokande  prevents 
from the shift of the allowed region 
to smaller mixings. Consequently, the combined  fit of the SNO and SK 
will also give the bound on the sterile fraction. 
Comparison of the bounds (\ref{etalimits2}) with  and (\ref{eta-snoG}) without  
the SK result allows to evaluate the role  of SuperKamiokande.\\

4). The Homestake  experiment alone is  
not sensitive to $\eta$.  However, it has some impact  since its result 
restrict the oscillation parameters.  
The $Ar$-production rate decreases with mixing angle 
and the agreement with the Homestake result becomes even better 
with  increase of $\eta$. Comparing the SNO and Homestake results 
we find  that in the best fit SNO point for $\eta = 0.6$ the rate 
equals $Q_{Ar} = 2.75$ SNU which is within $1\sigma$  of the  Homestake result. 
So, the SNO and Homestake data favor presence of 
sterile component. \\
 
Summarizing,  the bound on sterility appears as combined effect
of  several experiments.  
In the global analysis the constrain on $\eta$ 
comes from the SNO measurements of the spectrum which contains information 
about the NC and CC event rate from the one hand side 
and Gallium as well as the SK experiment from the other side. 
With increase of $\eta$,
SNO pushes the allowed region to smaller  $\tan^2 \theta$,
whereas  Gallium and SK experiments put lower bound on $\tan^2 \theta$.
For instance,  the best SNO fit point for $\eta = 0.6$ lies in the excluded
Gallium region and at the border of the SK excluded region.

\section{Future experiments and  bounds on $\eta$}

\subsection{$NC$ measurements at SNO}

SNO is taking data with much higher sensitivity to  the $NC$ 
current events, and  therefore  precision of measurements of 
$\Phi_{NC}$ will be significantly improved.
Clearly, distortion of the spectrum of the NC events will be signature of
the sterile neutrino conversion. Although if LMA is the dominant
solution, one does not expect observable effect. 
As far as total  rate is  concerned,  the parameters 
can be adjusted in such a way that  even  more precise data 
can be reproduced for large $\eta$.  Basically we will get the same shift of the 
allowed region as in  fig.~\ref{fig:chi2all} but the size of the region around the  
the best fit point will be smaller.  
For this reason no drastic improvements of the bound on sterility  is expected. 

\begin{figure}[ht]
\centering\leavevmode
\epsfxsize=.8\hsize
\epsfbox{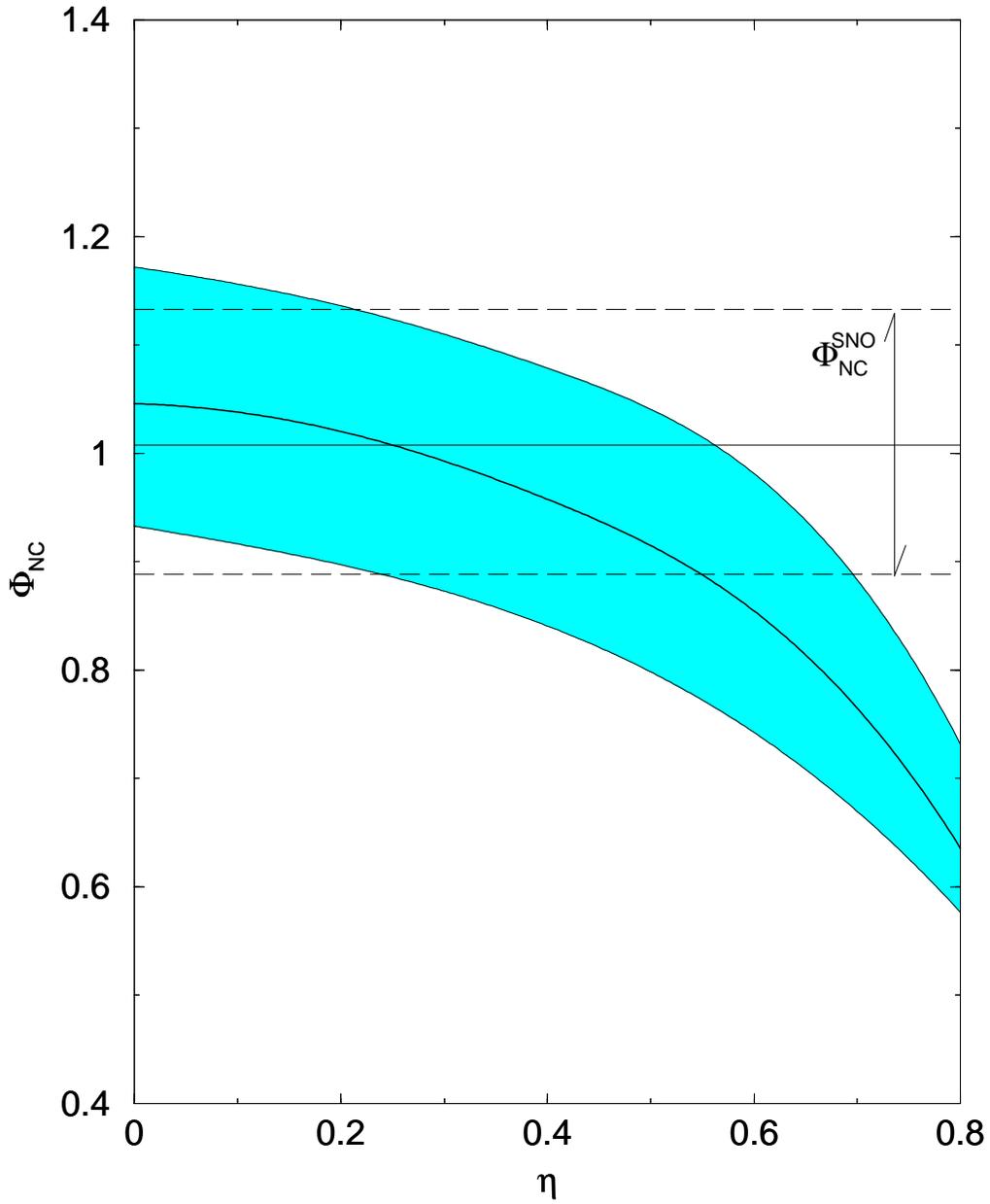}
\caption{The  predicted value of $\Phi_{NC}$ as a function of $\eta$. 
The band corresponds to the $1\sigma$ region. The horizontal lines show the present SNO 
result.}
\label{fig:ncpredictions}
\end{figure}

To evaluate the impact of future  $NC$-measurements at SNO we show in fig.~\ref{fig:ncpredictions} 
the $1\sigma$ region of predictions 
of  $\Phi_{NC}$  as a function of $\eta$ from the global fit. 
The results have been obtained in the following way. 
We use the global analysis of the solar neutrino data for different 
values of $\eta$, as is shown in fig.~\ref{fig:chi2all}. 
Then we calculate $\Phi_{NC}$ in the  best fit points (solid line)  as well as the spread
of predictions for $\Phi_{NC}$ within $1\sigma$ allowed region. 
Using the fig.~\ref{fig:ncpredictions},  we find that the present 
$1\sigma$ experimental result on $\Phi_{NC}$, 
gives the upper bound $\eta < 0.7$ which agrees with the 
$3\sigma$ bound from the global analysis.  Assuming three times smaller
error bars for the $NC$ measurements, $\Phi_{NC} = 1 \pm  0.03$,   we  get the bound $\eta < 0.5$  
which should be compared with present bound $\eta < 0.64$.

\subsection{KamLAND and sterile fraction}

The  signal  at KamLAND  is determined  by the vacuum oscillation survival 
probability, so that the rate of events is  
\be
N = \sum_i  \int dE \left(1 - \sin^2 2\theta_{12} \sin^2 
\frac{\Delta m^2_{12} L_i}{2E} \right) F_i \sigma A~,  
\label{n-rate}
\ee
where $F_i$ is the flux from $i$ reactor, $\sigma$ is the cross-section of 
$\bar \nu p \rightarrow e^+  n$ reaction, 
$A$ is the  energy resolution function.   
The sum is taken over all reactors contributing to the flux 
at Kamioka and the integral denotes schematically the integrations over
the neutrino energy, and the true energy of produced positron.  
This rate does not depend on $\eta$ or $f_B$. Being combined with 
the solar neutrino results it breaks degeneracy. 
KamLAND  will fix the oscillation parameters  $\Delta m^2_{12}$, $\tan^2 \theta_{12}$,   
and consequently,  will allow to predict $P_{ee}$ which describes the 
solar neutrino conversion. 

Let us evaluate how combined analysis of the solar data and KamLAND 
can improve the bound on  sterile component. 

Let us introduce the suppression factor of total event rate  
above certain threshold: 
\be
R_{KL}(\Delta m^2, \tan^2 \theta) \equiv \frac{N(\Delta m^2, \tan^2 \theta)}{N_0}~,
\ee
where $N$ is given in (\ref{n-rate}) and $N_0$ is the event rate  in absence of 
oscillations.  

Since  predictions for KamLAND do not depend on the state  to which the electron neutrino 
converts, we can use results of calculations of $R_{KL}$ 
for the pure active case~\cite{us:msw}.
In fig. 8 of paper \cite{us:msw} we have presented lines of constant $R_{KL}$ in
$(\Delta m^2, \tan^2 \theta)$ plane.  In the best fit points we find $R_{KL}(0) = 0.66$,  
$R_{KL}(0.2) = 0.63$, $R_{KL}(0.4) = 0.55$,   $R_{KL}(0.6) = 0.52$.  
Due to decrease of $\Delta m^2$  in the best fit points, the  
expected value of the suppression in the KamLAND experiment becomes stronger. 

\begin{figure}[ht]
\centering\leavevmode
\epsfxsize=.8\hsize
\epsfbox{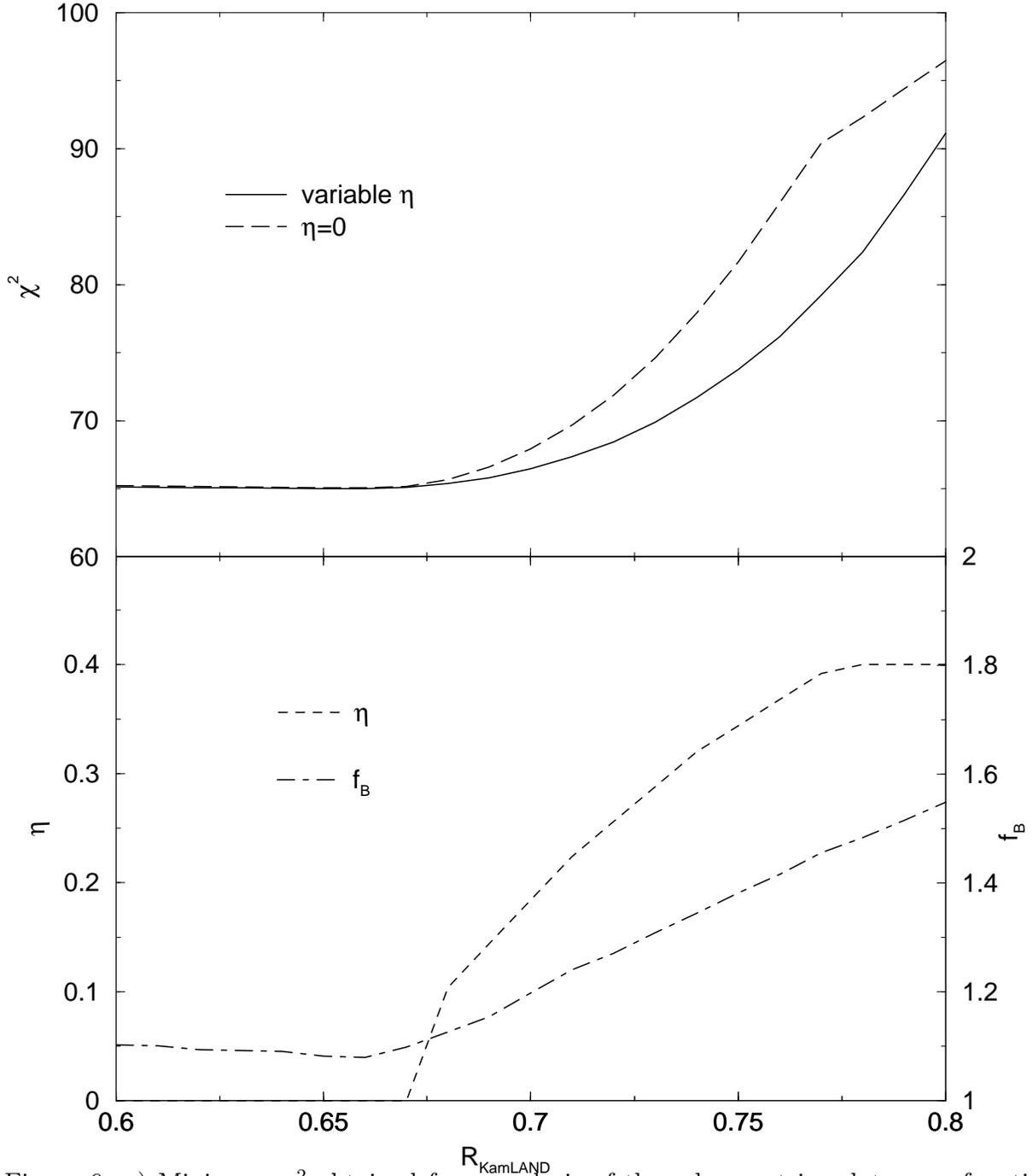}
\caption{ a) Minimum $\chi^2$ obtained from analysis of  the solar neutrino data
as a function of the KamLAND rate, $R_{KL}$, 
$\eta=0$ (dashed line) and  for $\eta$ taken as a free parameter (solid line).  
b) Value of $\eta$ (solid line and left axis) and $f_B$ (dashed line and right axis) 
that minimize the $\chi^2$ presented in the panel a).}
\label{fig:rkl}
\end{figure}

To quantify impact of the event rate measurements at KamLAND  
we have constructed the $\chi^2 (\Delta m^2, \tan^2 \theta, f_B, \eta)$ 
as the function $R_{KL}$. Practically we impose the following condition  
$R_{KL}(\Delta m^2, \tan^2 \theta) = const$ which implies  relation 
between $\Delta m^2$ and  $\tan^2 \theta$. So, for a given value  $R_{KL}$ the 
$\chi^2$ is the function  of $\Delta m^2$ or  $\tan^2 \theta$ only:
\be
\chi^2 = \chi^2 (\tan^2 \theta, R_{KL}, f_B, \eta). 
\ee
For each value of $R_{KL}$ we minimize $\chi^2$ with respects 
to $\tan^2 \theta, f_B, \eta$ thus finding 
$\chi^2_{\theta, f_B, \eta}(R_{KL})$. This function 
is shown in the upper panel of fig.~\ref{fig:rkl}. In the bottom panel  
the  corresponding values of $f_B$ and  $\eta$, which minimize $\chi^2$,  
are presented. For comparison in the upper panel we show also the function  
$\chi^2_{\theta, f_B}(R_{KL})$ for  $\eta = 0$.

The following remarks are in order. 

For $R_{KL} < 0.67$ the quality of the fit does not depend on $R_{KL}$. 
It corresponds to $\eta = 0$ and $f_B \approx 1.05$.  
For $R_{KL} > 0.67$, $\chi^2$  increases with $R_{KL}$. 
Simultaneously the best fit values of $\eta$ and $f_B$ increase. 
At $3\sigma$ level ($\chi^2 = 74$) we find 
maximal value of   $R_{KL}$ and corresponding values of $\eta$ and  $f_B$
\begin{equation} 
R_{KL} \sim 0.75, ~~~~\eta = 0.35, ~~~~ f_B = 1.4.  
\end{equation}
For $\eta = 0$ the value  $R_{KL} \sim 0.725$ is allowed.  So, the  
introduction of sterile neutrino in the single $\Delta m^2$ approach 
does not lead to  significant increase of maximal allowed value of $R_{KL}$. 
The goodness of the fit decreases sharply: {\it e.g.} for $R_{KL} = 0.8$ we get    
$\Delta \chi^2 = 25$ even in presence of sterile component.

\begin{figure}[ht]
\centering\leavevmode
\epsfxsize=.8\hsize
\epsfbox{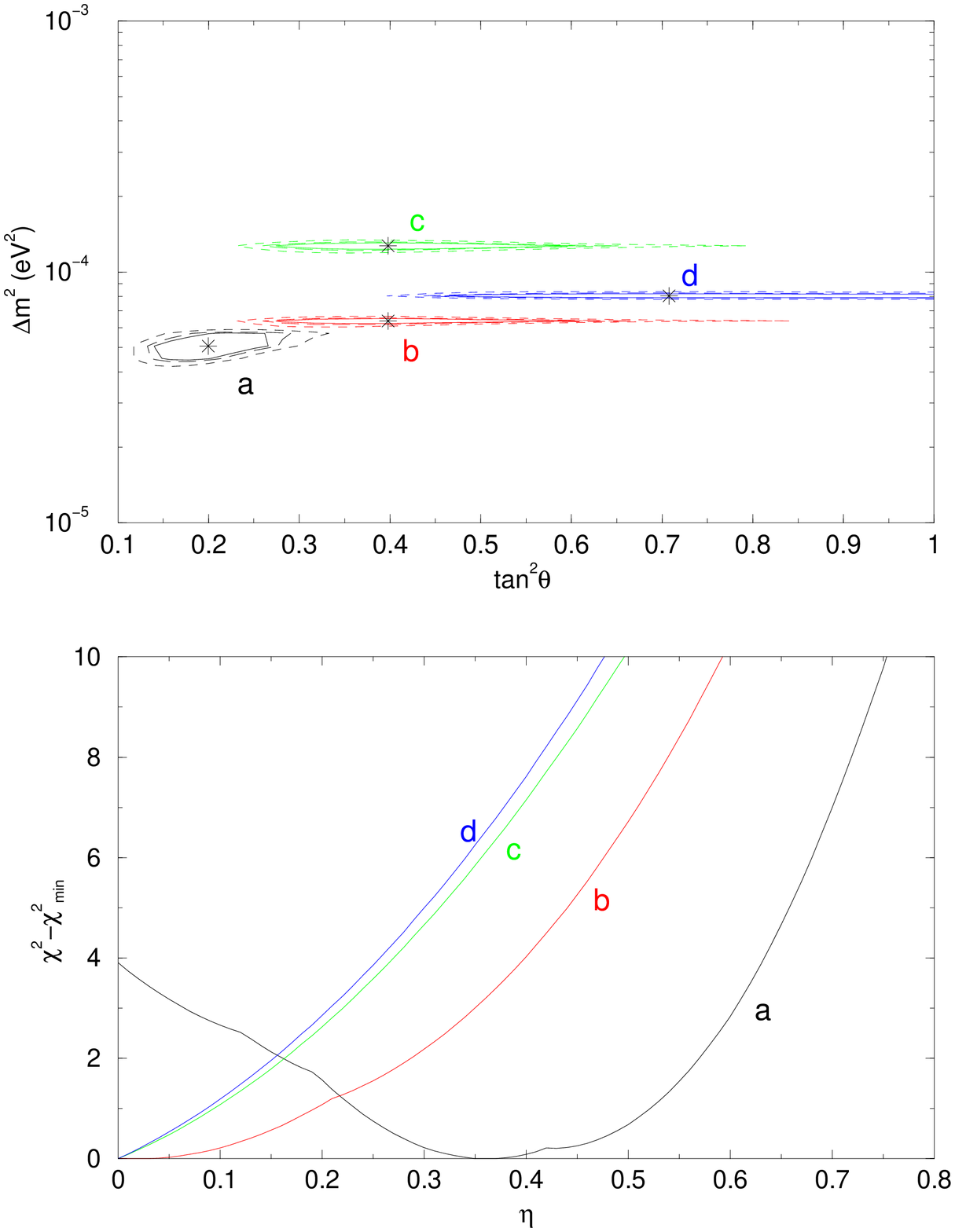}
\caption{a) The  $1 \sigma, 2\sigma, 3\sigma$ allowed regions from the analysis of 
simulated KamLAND data.  
True values for the neutrino parameters that are indicated by a
stars.  b) The dependence of the $\chi^2$  on $\eta$ for
these points  from a combined analysis of the simulated KamLAND  and 
the solar neutrino results.}
\label{fig:chi2kaml}
\end{figure}

Let us analyze how the KamLAND spectral measurements  can help in  constraining   $\eta$. 
We have simulated the KamLAND spectral data for  4 different 
points a), b), c), d)  in the $\Delta m^2- \tan^2 \theta$ plane 
(marked by stars in fig.~\ref{fig:chi2kaml}a)). 
In simulations we use statistics which can be collected during 3 years in 600 tons   
above the energy threshold for visible energy $E_{vis}=2.6$ MeV.  
In absence of oscillations this would correspond to 1170 events.  
We assumed a $4\%$ uncertainty in the overall normalization
and a $2\%$ uncertainty in the energy  scale.  
We  have simulated the energy spectrum (for each point) using 
12 bins of $0.5$ MeV size in the visible energy. 

We perform  $\chi^2$ analysis of simulated  spectra  
assuming  a full correlation of systematics uncertainties. We minimize   
\begin{equation}
\chi^2_{KL}=\sum_{i=1,12}\sum_{j=1,12}R_i\sigma^{-2}_{i,j}R_j  
\end{equation}
and construct contours of constant $1\sigma$, $2\sigma$, 
$3\sigma$ confidence level  around selected points in the $\Delta m^2,  \tan^2 \theta$ plane 
(see fig.~\ref{fig:chi2kaml}a). 
Notice that our allowed regions are somehow larger than those obtained 
in ~\cite{bargerMW,GP,milano}, since we include  systematic errors and also use higher 
threshold (smaller number of events). In agreement with results of calculations 
of other groups~\cite{bargerMW,GP,milano} we find  that KamLAND will be able  to measure 
$\Delta m^2$ rather precisely   (if $\Delta m^2 < 2 \cdot 10^{-4}$ eV$^2$). 
At the same time determination of the mixing angle will not be very accurate. 
For the present best fit point $(b)$  we get from the figure  
$\tan^2 \theta = 0.28 - 0.60$ at the $1\sigma$ level, and 
$\tan^2 \theta = 0.25 - 0.85 $ at the $3\sigma$ level. 

Then we have performed combined analysis of the existing solar and
simulated KamLAND data (notice that by the time KamLAND will have 3
years statistics new solar neutrino data will appear). For each
selected point of the oscillation parameters in
fig.~\ref{fig:chi2kaml}a we calculate a global $\chi^2_{tot}$, which
includes contributions from the solar neutrino analysis,
$\chi^2_{Sun}$, and the KamLAND, $\chi^2_{KL},$:
\begin{equation}
\chi^2_{tot} = \chi^2_{KL}(\tan^2\theta,\Delta m^2) + 
       \chi^2_{Sun}(\tan^2\theta,\Delta m^2,\eta,f_B) \,\,.
\end{equation}
For each value of $\eta$ we minimize  $\chi^2_{tot}$ with respect to
$\Delta m^2$, $\tan^2\theta$ and $f_B$. Results of this
minimization as a function of $\eta$  are presented in
fig.~\ref{fig:chi2kaml}b.  Let us comment on results for four selected points. 

The point $b)$ is the best fit point of the global fit 
which corresponds to $\eta = 0$. 
As follows from the figure,  
\begin{equation}
\eta < 0.19~(0.56)~~~ 1 \sigma~ (3\sigma), 
\end{equation}
thus improving the bound (\ref{etalimits2}) from  the solar data analysis only.   
  
The points c) and d) are at larger $\Delta m^2$ and $\tan^2 \theta_{12}$ as compared with b) 
thus  further  removed from the region 
preferred in the case of non-zero sterile component 
(small $\Delta m^2$ and $\tan^2 \theta_{12}$). If 
KamLAND  selects these points the bound on sterile 
component will be stronger than in a): 
$\eta < 0.09~(0.45)$ at $1\sigma~(3\sigma)$. 

The point a) is in the region preferable by 
sterile component.  Selecting this point KamLAND 
will favor non-zero $\eta$. From the lower panel 
we get: $\eta = 0.36^{+ 0.16}_{-0.12}$ at $1\sigma$ and 
$\eta > 0$ at $2\sigma$ level. The $1\sigma$ lower bound is $\eta > 0.24$.

\begin{figure}[ht]
\centering\leavevmode
\epsfxsize=.8\hsize
\epsfbox{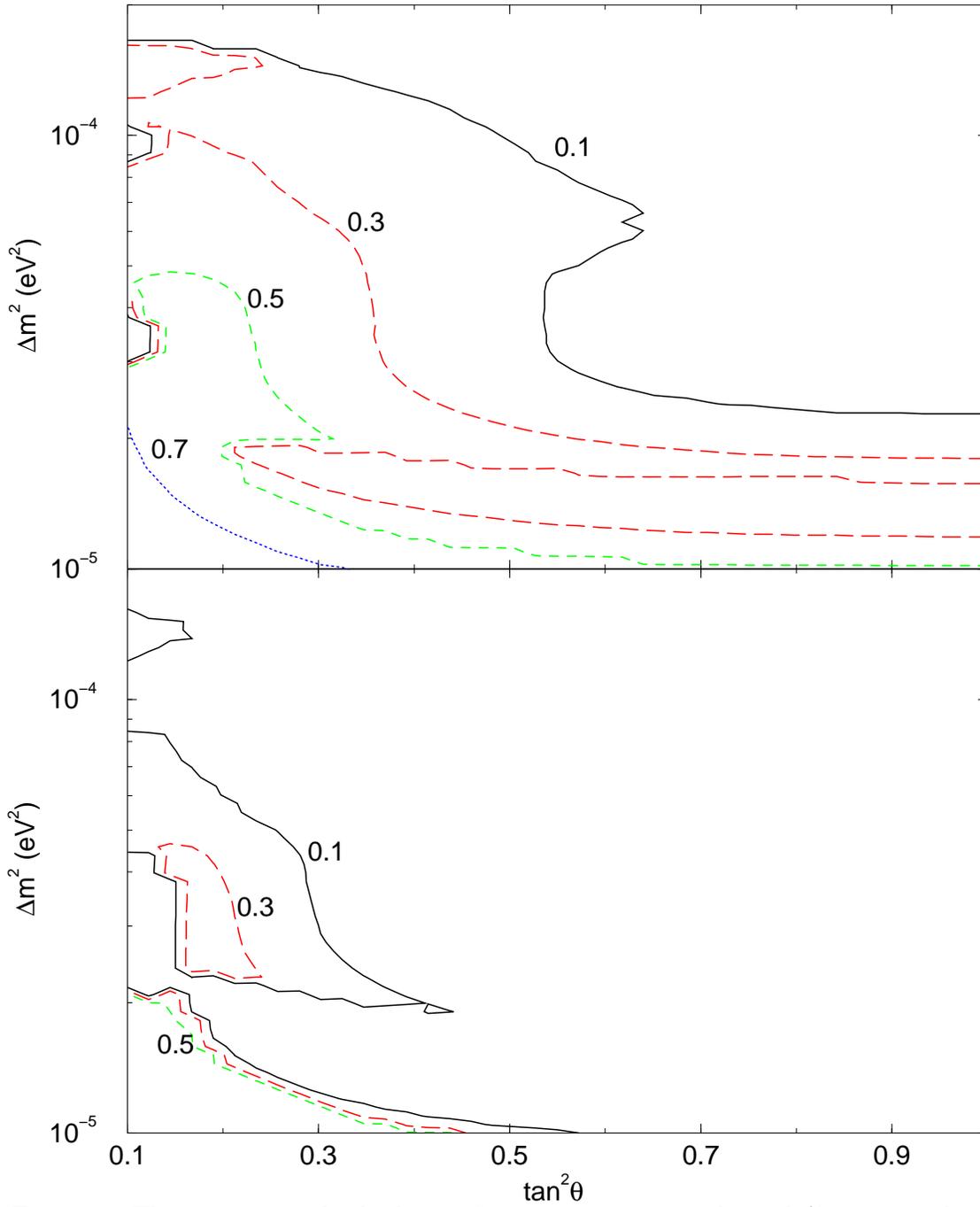}
\caption{The upper panel: the lines of constant 
1$\sigma$ upper bound (figures at the curves) on 
$\eta$ which can be obtained from  from  combined analysis of simulated KamLAND and 
solar neutrino results. 
The lower panel: the same as in the upper panel but for 
1$\sigma$ lower bound on $\eta$.}
\label{fig:etamax_kl}
\end{figure}

For the  points a) - d) we get the upper bounds on $\eta$: 
$\eta < 0.08$ (c and d), $0.19$ (b) and $0.52$ (a).   
We have performed similar analysis for all the points of the 
oscillation parameter space. We show in fig.~\ref{fig:etamax_kl} 
the results as lines of constant upper bounds in the oscillation parameter plane. 
As a tendency, the bound becomes weaker with decrease of 
mixing and $\Delta m^2$. 
For sufficiently small $\tan^2\theta$ the best fit corresponds to 
a non-zero value of $\eta$ and the lower bound on $\eta$ appears. 
In the lower panel of fig.~\ref{fig:etamax_kl} 
we show such a  lower limit for all points in 
$(\Delta m^2,\tan^2\theta)$ plane.\\

\section{Beyond single $\Delta m^2$ context}

The results described in the previous sections 
can change if the electron  neutrino mixes in the other mass eigenstates 
beyond the solar pair. Now more than one $\Delta m^2$ is involved, and   
effect depends on specific values of additional $\Delta m^2$.   

In what follows we will assume that mixing of $\nu_e$ in the states 
$\nu_{3}$ and $\nu_4$ is small. Then if additional $\Delta m^2$ are outside 
the range of the MSW enhancement ({\it e.g.} $\Delta m^2_{14}$ is large)  
the effect of new $\nu_e$ mixing is small being suppressed by the  mixing itself. 
If, however, $\Delta m^2_{14}$ is in the range of the MSW enhancement, 
the effect of additional mixing can be large. We will consider both possibilities.

\subsection{Large additional $\Delta m^2$}

Suppose  $\nu_e$ has non-zero admixture in $\nu_3$ characterized by 
by $U_{e3} = s_{13}$.  (Recall that  $\nu_3$ is separated from the solar neutrino pair 
by the mass gap which corresponds to the atmospheric neutrino split.) 
We assume  that $\nu_e$-admixture  in $\nu_4$ is negligible so that the  
mass of $\nu_4$  can be  arbitrary within the allowed range by non-solar neutrino 
experiment. We assume also  that in the limit $s_{13} \rightarrow 0$, the
state $\nu_3$ coincides with  $\nu_a'$ - the 
combination of $\nu_{\mu}$ and $\nu_{\tau}$, 
which is orthogonal to $\nu_a$ (\ref{aaa}). In turn, $\nu_4$ is the
combination of $\nu_s$ and $\nu_a$. 
As a result, we get the following mixing pattern: 
\begin{eqnarray} 
\nu_1 &=& \cos\theta_{12}(c_{13}\nu_e  - s_{13}\nu_a') -
\sin\theta_{12} \nu_x, \nonumber\\
\nu_2 &=& \sin\theta_{12}(c_{13}\nu_e  - s_{13}\nu_a') + 
\cos\theta_{12} \nu_x,\nonumber\\
\nu_3 &=& c_{13}\nu_a' + s_{13}\nu_e, \nonumber\\ 
\nu_4 &=& \nu_x', 
\label{mix4}
\end{eqnarray} 
where $\nu_x$ is defined in (\ref{nux}) and $\nu_x'$ is the orthogonal 
to $\nu_x$ combination of $\nu_s$ and $\nu_a$. 
The mixing (\ref{mix4}) differs from the generic case  since 
$\nu_e$ and  $\nu_a'$ do not mix in $\nu_4$, and $\nu_a$ and $\nu_s$ 
do not mix in $\nu_3$. 

The dynamics of propagation is similar to that of the case of three
neutrinos. Indeed, let us represent $\nu_e$ in the following form: 
\be
\nu_e = c_{13}\nu_e'  + s_{13} \nu_3~,  
\label{nue}
\ee 
where $\nu_e' =  \cos\theta_{12} \nu_1 + \sin\theta_{12} \nu_2~$ 
coincides with $\nu_x$, as can be easily checked. For the energies of solar neutrinos 
the matter effect on $s_{13}$  is negligible.  
Furthermore,  the oscillations of solar neutrinos 
related to $\nu_3$ are averaged out. 
At the same time, $\nu_e'$ and $\nu_x$ form standard two neutrino 
system with vacuum mixing and mater effects. 

Using these features of dynamics, as well as  projection of original
flavor states onto $\nu_e', \nu_x, \nu_3$ 
one gets immediately the conversion probabilities: 
\begin{eqnarray}
P_{ee} &=& c_{13}^4 \tilde{P}_{ee} + s_{13}^4 \nonumber\\
P_{es} &=& c_{13}^2 \eta(1 -   \tilde{P}_{ee}) \nonumber\\
P_{ea} &=& c_{13}^2(1 -  \eta) (1 -   \tilde{P}_{ee}) + 
s_{13}^2 c_{13}^2  (1 +  \tilde{P}_{ee}),   
\label{probab}
\end{eqnarray} 
where $\tilde{P}_{ee}$  is the
$\nu_e'$-survival probability in the
$\nu_e'-  \nu_x$ system: $\tilde{P}_{ee} = \tilde{P}_{ee}(\theta_{12},
\Delta m^2_{12}, \tilde{V})$ and 
$\tilde{V} = c_{13}^2 V_e - V_{\mu}(1 - \eta)$. 
As in the previous case,  the parameter $\eta$ 
determines fraction of $\nu_s$ in the state $\nu_x$ to which
$\nu_e'$ converts, and also it gives total fraction of the $\nu_s$
in the solar pair (\ref{norm}).

Using (\ref{probab}), we get  fluxes (in the units of the SSM flux) 
measurable by different reactions:
\be
\Phi_{CC} = f_B[c_{13}^4 \tilde{P}_{ee} + s_{13}^4] ,
\label{cc-flux}
\ee
\be
\Phi_{NC}  = f_B[1 - (1 - \tilde{P}_{ee})\eta c_{13}^2] , 
\label{nc-flux}
\ee
\be
\Phi_{ES} = f_B[c_{13}^4 \tilde{P}_{ee}  + s_{13}^4
+ r c_{13}^2 (1 - \tilde{P}_{ee})(1 - \eta) + r s_{13}^2(1 +
\tilde{P}_{ee})].
\label{esflux}
\ee
Comparing with Eq. (\ref{eq:RSNOCC},\ref{eq:RSNONC}, \ref{eq:RSK}) we find 
that corrections are indeed of the order $s_{13}^2$. 
For the allowed values of $s_{13}^2$ these corrections have small impact 
on the data fit, as one can see from our analysis in \cite{us:msw}. 

The fluxes $\Phi_{CC}$, $\Phi_{NC}$,  $\Phi_{ES}$ 
depend on two combinations of parameters, as in Eq. (\ref{fluxeq}),  
with substitution $x \rightarrow x'$, $y \rightarrow y'$, where   
\be
x' = c_{13}^4 x, ~~~~y' = c_{13}^2 + s_{13}^2 (x + f_B),  
\ee
and $x$ and $y$ are defined in (\ref{twocomb}).  
So, the problem of degeneracy is not resolved with $s_{13}$ corrections. 
Notice also that the fluxes in (\ref{cc-flux} - \ref{esflux}) 
satisfy the relation (\ref{sum}).

The KamLAND signal is determined by the survival probability 
\be
P_{ee} =  c_{13}^4 {P}_{2} + s_{13}^4, 
\ee
where $P_{2}$ is the  two neutrino vacuum oscillation probability 
determined by $\Delta m^2_{12}$ and $\theta_{12}$. 
The factor $c_{13}^4$ leads to additional suppression of the KamLAND signal and shift of 
the allowed regions to smaller $\theta_{12}$ 
(see detailed analysis in the $3\nu$ context in \cite{ue3}). 

\subsection{Small additional $\Delta m^2$}
\label{small}

Let us consider the case when even small $\nu_e$ mixing beyond the 
solar pair can produce a strong effect.  Suppose that,  

(i)  $\nu_e$ mixes with $\nu_x$ in the 
solar pair, $\nu_1$ and $\nu_2$,   
which has $\Delta m_{12}^2$ and $\tan^2 \theta_{12}$ in the LMA region,   
as in (\ref{mix4}); 

(ii)  the state $\nu_4 \approx \nu_x'$ (see (\ref{mix4})) 
is heavier than $\nu_1$ 
and the   mass splitting is  $\Delta m_{41}^2 \sim 10^{-6}$ eV$^2$. 
(The mass $m_1$ can be very small or even zero). 
Furthermore, $\nu_e$ has small admixture in $\nu_4$: 
$\tan^2 \theta_{14}  \sim 10^{- 3}$, 
so that $\nu_e - \nu_x'$ oscillation parameters are in the SMA region.  

(iii) the admixture of $\nu_e$ in the $\nu_3$ is zero (for simplicity). 

Then, similarly to (\ref{mix4})  we can write the mixing  as 
\begin{eqnarray}
\nu_1 &=& \cos\theta_{12}(c_{14}\nu_e  - s_{14}\nu_x') -
\sin\theta_{12} \nu_x, \nonumber\\
\nu_2 &=& \sin\theta_{12}(c_{14}\nu_e  - s_{14}\nu_x') +
\cos\theta_{12} \nu_x,\nonumber\\
\nu_3 &=& c_{13} \nu_a', \nonumber\\
\nu_4 &=&  c_{14}\nu_x' + s_{14}\nu_e.
\label{mix4small}
\end{eqnarray}
Here $\nu_x'$ is the combination of $\nu_a$  and $\nu_s$ orthogonal to 
$\nu_x$ (\ref{nux}), and $\nu_a'$ is the combination of $\nu_{\mu}$  and $\nu_{\tau}$ 
orthogonal to $\nu_a$ (\ref{aaa}).

The system has two resonances associated with 
$\Delta m_{12}^2$ and $\Delta m_{41}^2$. 
Let us estimate qualitatively the conversion effect in different energy ranges.

In the first resonance related to $\Delta m_{12}^2$ the conversion is completely adiabatic. 
So, at the high energies, $E > 7  - 8$ MeV,  
where  crossing of the resonance  occurs 
the survival probability equals $P_{ee}^H \approx \sin^2\theta_{12}$. 
At low energies, $E < 0.5$ MeV,  (no 12-resonance crossing) 
effect driven by  $\Delta m_{12}^2$  is  reduced to averaged 
vacuum oscillations. 
The result of conversion  is similar to the one in usual three neutrino 
system: 
\be
P_{ee}^L = c_{12}^4 P_{2}' + s_{12}^4,
\label{pprime}
\ee
where $P_{2}'$ is the survival probability in the resonance related to  
$\Delta m_{14}^2$. It has a typical form of the survival probability 
of the SMA solution and characterized by $\Delta m_{14}^2$ and 
mixing parameter $\tan^2\theta = \tan^2 \theta_{14}/ \cos^2\theta_{12}$. 
Appearance of $P_{2}'$ in this expression leads 
to an additional suppression of the $pp$-neutrino flux. 

In the intermediate energy range,  $E = (0.5 -  7)$ MeV, 
the conversion driven by  $\Delta m_{14}^2$ 
will lead to flattening of the adiabatic edge 
of the suppression pit due to $\Delta m_{12}^2$.

\subsection{KamLAND result and LMA} 

In the  best fit point of the LMA  region 
we  predict  $R_{KL} = 0.66$ for   zero 
$\eta$~\cite{us:msw}. At $3\sigma$ level $R_{KL} < 0.73$. In presence of 
the sterile component this bound can be slightly relaxed: 
$R_{KL} < 0.76$. For other solutions of the solar neutrino problem 
we predict $R_{KL} \approx 1.0$ if $s_{13} = 0$. 
For non-zero $s_{13}$ the oscillations driven by 
$\Delta m^2_{atm}$ will lead the averaged oscillation result at 
KamLAND: $R_{KL} = 1 - 0.5 \sin^2 2\theta_{13}$. For maximal allowed  
values of $s_{13}$ we get $R_{KL} = 0.90 - 0.95$ depending on $\Delta m^2_{atm}$. 
What if $R_{KL} = 0.76 - 0.90$ will be found? 

The following comments are in order: 

If solution of the solar neutrino problem is  in the  LMA region, 
a  strong suppression  is expected in KamLAND.
In this case, introduction of additional sterile states 
will not help in the  single $\Delta m^2$ context, 
as we have found  in the Sect. \ref{glo}.  

Let us discuss what happens if two or more  $\Delta m^2$ contributes. 

Oscillation parameters can be taken beyond the 
LMA region in such a way that suppression in the KamLAND experiment is  weaker.
In this case, however,  the conversion of  solar neutrinos 
will not describe data  in the two neutrino context. 
Then the parameters (mass, flavor mixing) of the $\nu_4$ can be selected in such a way 
that conversion driven by  $\nu_4$ will not change KamLAND result but correct 
description of the solar data. 

Such a possibility can be realized in the scheme described in the 
previous section \ref{small} with  $\Delta m_{12}^2 \sim 6 \cdot 10^{-5}$ eV$^2$. 
Indeed the KamLAND result is determined by the  
oscillations driven by $\Delta m_{12}^2$ and $\theta_{12}$. 
For $\tan^2 \theta_{12} \sim 0.10 - 0.15$ we get $R_{KL} = 0.80  - 0.85$.  

As far as the solar neutrino signal is concerned,   
the high energy signal will be suppressed by $\sin^2 \theta$, so that 
large boron neutrino flux: $f_B \sim 1.6$ is required to explain the 
SNO and SK results.   
According to our previous consideration,   in the 
single $\Delta m^2$ context this point would be  excluded by the Gallium experiment 
(large $Q_{Ge}$) and by SK (distortion of spectrum). 
The effect of $\Delta m_{14}^2$ changes a situation: 
appearance of $P_{2}'$ in  expression for 
probability at low energies (\ref{pprime}) leads 
to an additional suppression of the $pp$-neutrino flux, and consequently, 
$Q_{Ge}$ rate. In the intermediate energy range,  $E = (0.5 -  7)$ MeV, 
the conversion driven by  $\Delta m_{14}^2$ 
will relax the SuperKamiokande  
lower bound on mixing which follows from absence of spectrum distortion.  
The detailed analysis of this possibility will be given elsewhere~\cite{future}. \\

Another possibility to reconcile  an intermediate  suppression of 
signal in KamLAND  and explanation of the solar neutrino data is 
to split two problems. 
Suppose the solar pair has $\Delta m_{12}^2$  in the LOW or VO regions, 
so that it does not produce any effect in KamLAND. 
The state $\nu_4$ which consists predominantly of the sterile component  
has the mass split: $\Delta m_{14}^2 > 10^{-4}$ eV$^2$. 
Then the suppression of the KamLAND signal is determined by admixture 
of $\nu_e$ in $\nu_4$: $R_{KL} = 0.5 \sin^2 2\theta_{14}$. 
For instance, $R_{KL} = 0.85$ can be achieved 
at $\sin^2 \theta_{14} = 0.09$. 
Effect in solar neutrinos will be determined by 
$P_{ee}^{sun} = c_{14}^4 P_{ee} + s_{14}^4$, where $P_{ee}$ is 
the survival probability for the conversion driven by $\Delta m_{12}^2$.  
Additional factor $c_{14}^4 \sim 0.8$ can even improve the 
fit of the data for LOW and SMA~\cite{us:msw}.

\section{Conclusions}

1. We have studied properties of mixing of the sterile and active 
neutrinos. In the single  $\Delta m^2$ context (when electron neutrino 
is  present in two mass eigenstates only)  the problem is reduced
precisely to  
two neutrino  problem with mixing of $\nu_e$ with $\nu_x$ in the solar
pair ($\nu_x$ is  the  combination of the active and sterile neutrinos).

\noindent
2. We have performed the  global fit of the  solar neutrino data 
in the  single $\Delta m^2$ context, treating  
the boron neutrino flux  as  a free parameter. 
The best fit corresponds to zero admixture of sterile component, 
$\eta = 0$. We get the upper bound 
$\eta < 0.26~ (0.64)$ at $1\sigma ~~(3\sigma)$. 
With increase of $\eta$  the best fit point shifts to smaller 
$\Delta m^2$ and $\tan^2 \theta_{12}$.

\noindent
3. Due to degeneracy of parameters, no one single 
observable gives the bound on $\eta$. The bound on sterility appears as an interplay of the SNO results 
whose allowed region of the oscillation parameters shifts to 
smaller mixings and $\Delta m^2$ with increase of $\eta$ and 
Gallium as well as SK results which restrict the mixing and 
$\Delta m^2$  from below.  

\noindent
4. Further more precise measurements of the $NC$ event rate at SNO will strengthen the 
bound: $\eta < 0.5$ ($3\sigma$). 

\noindent
5. Implications of   KamLAND depend on specific 
results of KamLAND measurements. If KamLAND result (rate, spectrum) corresponds 
to the best fit point of the solar neutrino analysis,  
then after 3 years of the KamLAND operation the limit will be strengthened  
down to $\eta < 0.19~~(0.55)$ at $1\sigma ~~(3\sigma)$. 
The limit will be stronger if the KamLAND best fit point will be at larger 
$\tan^2 \theta_{12}$  and $\Delta m^2$ than the solar neutrino data
give at present. 
In contrast, if  KamLAND selects smaller $\tan^2 \theta_{12}$ and $\Delta m^2$, 
the limit will be weaker, and moreover, the data may prefer non-zero 
sterile admixture.

\noindent
6. The presence of sterile neutrinos (in the single $\Delta m^2$ context) 
relaxes the upper bound on the predicted rate $R_{KL}$ 
only weakly from 0.73 to 0.76 $(3\sigma)$. 
Larger values of $R_{KL}$ can be reached in the two $\Delta m^2$ 
context. 

\noindent
7. If electron neutrino mixes also in other mass eigenstates, the 
effects depend substantially on values of other $\Delta m^2$. 

If additional $\Delta m^2_{14} \gg \Delta m^2_{12}$ the small corrections appear 
to  formulas with a single $\Delta m^2$  which are proportional to the mixing of 
$\nu_e$ in additional mass eigenstates, {\it e.g.}, $s_{13}^2$. 
If $\Delta m^2$, is in the MSW  range an additional 
mixing can be resonantly enhanced in matter 
producing much larger effect. In this case intermediate values of $R_{KL}$ 
can be obtained.

\newpage

\newpage

\newpage

\newpage

\newpage

\newpage

\newpage

\newpage



\end{document}